%% file: ms_apjl.tex
\documentclass{emulateapj}

\newcommand\Rstar{\hbox{$R_*$}}
\newcommand\ro{\hbox{$r_o$}}

\newcommand\TX{\hbox{$T_X$}}

\newcommand\etal{\hbox{et~al.}}

\newcommand\Chandra{\hbox{\it Chandra}}

\newcommand\zoriA{\hbox{$\zeta\ ${\rm Ori A}}}

\newcommand\cygate{\hbox{Cyg OB2 No. 8a}}

\newcommand\doriA{\hbox{$\delta\ ${\rm Ori A}}}
\newcommand\tsco{\hbox{$\tau\ ${\rm Sco}}}
\newcommand\toriC{\hbox{$\theta^1$ Ori C}}

\newcommand\vinf{\hbox{$V_\infty$}}
\newcommand\ves{\hbox{$V_{es}$}}
\newcommand\VA{\hbox{$V_{Apo}$}}
\newcommand\VP{\hbox{$V_P$}}
\newcommand\VW{\hbox{$V_W$}}

\newcommand\kms{\hbox{km$\,$s$^{-1}$}}

\newcommand\Alfven{\hbox{Alfv\'{e}n}}

\slugcomment{Accepted for publication in {\it The Astrophysical Journal Letters} }
\shorttitle{OB Plasmoid Ejecta}
\shortauthors{Waldron \& Cassinelli}

\begin{document}

\title{Highly Accelerated Diamagnetic Plasmoids: A New X-ray Production Mechanism for OB
Stellar Winds \\ }
\author {Wayne L.~ Waldron\altaffilmark{1} and Joseph P.~ Cassinelli\altaffilmark{2}}
\altaffiltext{1}{Eureka Scientific, Inc., 2452 Delmer St., Oakland CA, 94602;
wwaldron@satx.rr.com}
\altaffiltext{2}{Dept. of Astronomy, University of Wisconsin-Madison, Madison, WI 53711;
cassinelli@astro.wisc.edu}

\begin{abstract}
The observed X-ray source temperature distributions in OB stellar winds, as determined from high
energy resolution $Chandra$ observations, show that the highest temperatures occur near the
star, and then steadily decrease outward through the wind. To explain this unexpected behavior,
we propose a shock model concept that utilizes a well-known magnetic propulsion mechanism;
{\it the surface ejection of ``diamagnetic plasmoids'' into a diverging external magnetic field}. 
This produces rapidly accelerating self-contained structures that plow through an ambient wind
and form bow shocks that generate a range in X-ray temperatures determined by the 
plasmoid-wind relative velocities.
The model free parameters are the plasmoid initial \Alfven\ speed, the initial plasma-$\beta$ of the
external medium, and the divergence rate of the external field. These are determined by fitting the
predicted bow shock temperatures with the observed OB supergiant X-ray temperature
distribution. 
We find that the initial external plasma-$\beta$ has a range between 0 and 2, and the assumed
radially-decreasing external magnetic field strength that scales as $r^{-S}$ has a value of $S$
lying between 2 and 3. Most importantly, the initial plasmoid \Alfven\ speed is found to be
well-constrained at a value of 0.6 \vinf, which appears to represent an upper limit for all normal
OB stars. This intriguing new limit on OB magnetic properties, as derived from $Chandra$
observations, emphasizes the need for further studies of magnetic propulsion mechanisms in these
stars.
\end{abstract}
\keywords {shock waves -- stars: early-type -- stars: magnetic fields -- stars: winds, outflows -- 
X-rays: stars -- X-rays: general}

\section{Introduction}
Waldron \& Cassinelli (2007, 2008; hereafter, WC07 and WC08) presented a detailed study of the
$Chandra$ HETGS data from 17 early-type stars (hereafter, OB stars). One of the more
surprising results is the evidence for a radially dependent X-ray temperature (\TX)
distribution that steadily decreases outward through the stellar wind from a high value near the
star. 
We suggest that this distribution may be related to stellar magnetic field effects, and present
a semi-empirical model for X-ray production in OB stars that incorporates a well known magnetic
ejection process.  

The plausibility of magnetic ejection is supported by the increasing number of magnetic field
detections on OB stars. There are now direct measurements of magnetic fields for three O-stars
(Donati \etal\ 2002, 2006a; Bouret \etal\ 2008) and five early B-stars (Donati \etal\ 2001, 2006b;
Neiner \etal\ 2003; Petit \etal\ 2008). The recent observation by Bouret \etal\ (2008) provides the
first detection of a magnetic field on an O supergiant, \zoriA\ (O9.7 Ib), with a field strength of
60 - 100 G. 
It also has a complex magnetic topology apparently similar to the magnetic field structure of 
\tsco\ (B0 V) that is interspersed with ``hole regions'' that may resemble the magnetospheric
structure of the Sun (Donati \etal\ 2006b). 
This \zoriA\ detection is of particular interest to us since magnetic field strengths similar to the
observed value were proposed to explain its high energy X-ray emission lines in $Einstein$ SSS
data (Cassinelli \& Swank 1983) and $Chandra$ HETGS data (Waldron \& Cassinelli 2001). 

Theoretical studies of magnetic fields in OB stars have primarily focused on studying origins;
dynamo processes (Charbonneau \& MacGregor 2001; MacGregor \& Cassinelli 2003; Mullan \&
MacDonald 2005), and remnant fossil fields (Ferrario \& Wickramasinghe 2005, 2006). In
addition, Maeder \etal\ (2008) has shown that rapid rotation can extend the thickness of the thin
outer convection zones in OB stars (Maeder 1980). At this early stage of studying OB surface
magnetic field structures we can only surmise that these field structures may likely be analogous
to solar emerging magnetic field regions (Zwaan 1985). Unlike recent research
in solar physics our intent is not to explore ways of producing coronal regions around these stars,
but rather to explain the observationally derived X-ray temperatures of hot shock fragments that
are dispersed throughout their stellar winds.

\section{The \TX\ Distribution in OB Stars} 
\begin{figure}
\includegraphics[width=7cm,angle=-90]{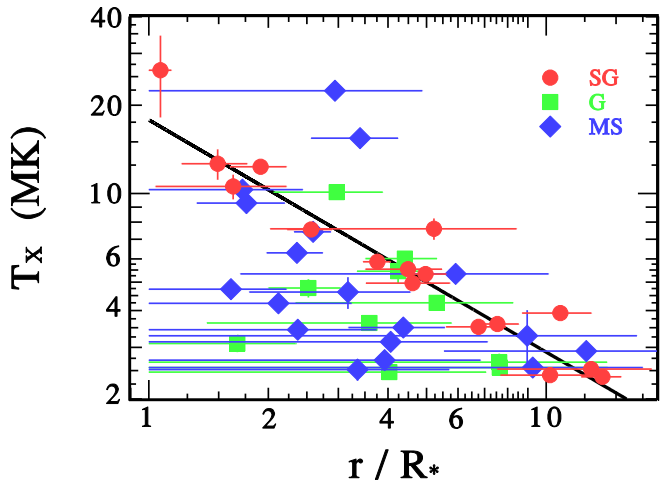}
\caption{Comparison of the OB supergiant power-law fit given by Eq. (\ref{eq:TPL}) (black line)
with the observed \TX\ for all OB supergiants (SG), giants (G), and main sequence (MS) stars as
determined by WC08.  The two data points significantly outside this bound are associated with
the peculiar MS star \toriC\ which has a very strong magnetic field of $\approx 1000$ Gauss
(Donati \etal\ 2002).
\label{fig:OBS}}
\end{figure}

At the core of our semi-empirical model is the observed radial decreasing source \TX\ distribution
found by WC07 (and updated in WC08) which is most noticeable for OB supergiants (see Fig.
\ref{fig:OBS}). This distribution is unexpected based on wind shock theory developed prior to
the launch of $Chandra$ (e.g., Feldmeier 1995). 
To analyze the significance of this \TX\ distribution we first obtain an empirical power-law fit
(error-weighted) to the WC08 data (neglecting the \doriA\ Ne IX data for reasons discussed by
WC07) given by 
\begin{equation}
\TX (r) ~ = ~ (18 \pm 2)~ {\left ( \frac {\Rstar} r \right )}^{0.79 \pm 0.05}~~~ MK
\label{eq:TPL} 
\end{equation}
This power-law fit shows that very high \TX\ are present near the star and we think this
expression will be useful for interpretations of the emission line results and for general modeling
efforts. As shown in Figure \ref{fig:OBS}, the lower luminosity class OB stars have
more complex \TX\ distributions since these stars have winds that are more optically thin to
X-rays, and there can be radially extended emission regions associated with any \TX\ value.
Nevertheless, Figure \ref{fig:OBS} shows that nearly all giant and main sequence stars have an
upper bound on \TX, and this upper bound is consistent with the supergiant power-law fit (Eq.
\ref{eq:TPL}). As discussed by WC07, clarifications of these observed \TX\ distributions should
lead to a better understanding of the overall X-ray emission process in OB stars, and in \S\ 3 we
present a model that provides a viable explanation of this radially-dependent \TX\ upper bound. 

\section{Plasmoid Model}
The primary goal of this paper is to find a process that can generate high X-ray temperatures at
small stellar radii ($\le 1.5$ \Rstar). In the wind-clump model of Howk \etal\ (2000) they realized
a need for high \TX\ at small radii and chose to consider the possibility of in-fall. Here we
consider an alternative method of producing large relative velocities at small radii by using an
early magnetic propulsion model suggested by Cargill \& Pneuman (1984; hereafter CP84). They
derived, in a straightforward way, the dynamics of isolated ``diamagnetic plasmoids'' that are
launched into a diverging external magnetic field. The plasmoids are produced by magnetic
reconnection events and are rapidly accelerated owing to the interaction of the plasmoid magnetic
moment with the external field (Schluter 1957; Parker 1957; Pneuman 1983). We picture these
plasmoids as being produced just above the photosphere, and due to the rapid acceleration, plow
through the overlying stellar wind that has drawn out the external magnetic field. The
plasmoid-wind interactions produce ``bow shock'' structures around the plasmoids that are likely
to be similar to those described by Cassinelli \etal\ (2008; hereafter CAS08), and the \TX\ will
depend on the relative velocities between these plasmoids and the wind. An advantage of bow
shock produced X-ray emission is that the emission measure is determined by the influx of the
ambient wind material. 

We neglect the effects of the line radiation force on these plasmoids because Abbott (1982) found
that the line driving is dominated by those spectral lines that have energies slightly larger than the
energy associated with the peak of the radiation field. Hence, for a parcel of gas that is in a higher
stage of ionization than expected from the radiation temperature, the line force will be greatly
reduced. The gas temperature of the plasmoid is likely to be higher than the surroundings because
the magnetic reconnection process should generate sufficient energy to increase the temperature
ionization state of the plasmoid (e.g., a gas temperature $>$ 0.2 MK should be sufficient to shift
the dominant ions beyond the radiation field peak of OB stars).  
The plasmoid is expected to remain hot since it will heat up as it accelerates as shown by CP84. 
In addition, the X-rays from the bow shock formed on the outer plasmoid surface will also
contribute to maintaining the higher level of ionization by the Auger effect as discussed CAS08.

The temperature of the X-ray source behind a shock depends on the relative velocity of the
incident material and the shock front. Consequently, to produce the high \TX\ at low radii ($<
1.5$ \Rstar) where the ambient wind velocity is low, these plasmoids must experience very rapid
accelerations to generate the required large relative velocities. As shown by CP84, the largest
accelerations can be achieved only for the case when the initial plasma-$\beta$ of the plasmoid,
$\beta_{po}$, is $<< 1$, where $\beta_{po} ~ = 8 \pi P_p (\ro) / {B_p (\ro) }^2$ is the ratio of
the initial plasmoid gas pressure to the initial magnetic pressure at the radius (\ro) where the
plasmoid is created.  
Using the momentum equation and the basic assumptions given by CP84, along with this
condition on $\beta_{po}$, one can obtain an analytic expression for the plasmoid velocity
(assuming the plasmoid velocity = 0 at \ro), 
\begin{equation}
{\VP (r) }^2 =  4 {\VA}^2 \left( 1 - \gamma (r) \left ( \frac {\ro} {r}
\right )^{S/2} \right) - {\ves}^2 \left( 1 - \frac {\ro}{r} \right)
\label{eq:VP} 
\end{equation}
where $\VA = B_p (\ro) / {(4 \pi \rho_p (\ro))}^{ 1 / 2}$ is the initial plasmoid \Alfven\ speed,
$\rho_p (\ro)$ = initial plasmoid mass density, \ves\ is the stellar effective escape speed, and $S$
is the exponent of the assumed radially-dependent external magnetic field, $B_e (r) = B_e
(\ro)~(\ro / r)^S$.  

\begin{figure}
\includegraphics[width=7.5cm,angle=180]{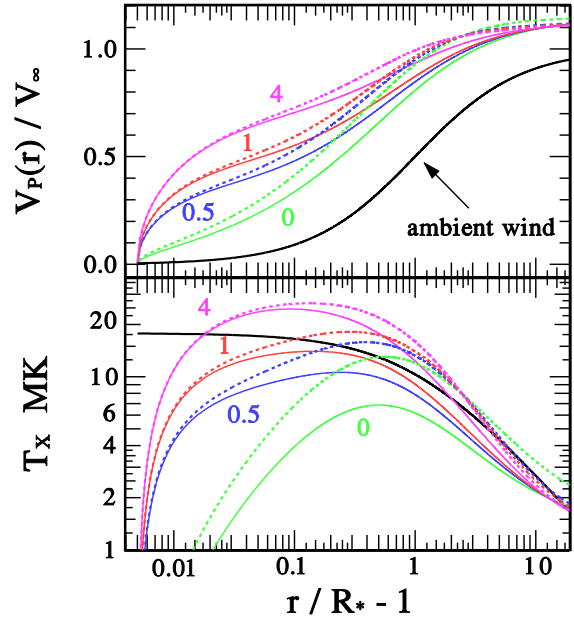}
\caption{Same as Figure 2 but for the parameter study of $\beta_{eo} = 0, ~ 0.5, ~ 1, ~ 4$, using
a fixed value of $\VA / \vinf = 0.6$. The radius axis has been re-scaled to magnify the details at
small radii where $\beta_{eo}$ effects are most prominent. \label{fig:BETA}}
\end{figure}

Equation (\ref{eq:VP}) is more general than the one given by CP84 due to the presence of
$\gamma (r)$ which arise from the external gas pressure gradient contribution to the plasmoid
acceleration. By allowing for a non-zero $\beta_{eo}$ we derived equation (\ref{eq:VP}) by
starting with CP84 equation (2.16), and found that $\gamma (r)$ is 
\begin{equation}
\gamma (r) = {\left( \frac {1 ~ + ~ \beta_e (r) } {1 ~ + ~ \beta_{eo}} \right ) }^{1/4} 
\label{eq:VPG} 
\end{equation}
where $\beta_e (r) = 8 \pi P_e (r) / {B_e (r)}^2$ is the radially-dependent external 
plasma-$\beta$, and $\beta_{eo} = \beta_e (\ro)$. In the limit of $\beta_e (r) \rightarrow 0$
(i.e., $\beta_{eo} \rightarrow 0$ and $\gamma \rightarrow 1$), equation (\ref{eq:VP}) reduces to
equation (3.2) of CP84. 
In this initial study we assume that the external temperature is isothermal and equal to the stellar
effective temperature and the overlying wind is spherical symmetric which are common
assumptions used in the development of basic wind models (e.g., Lamers \& Cassinelli 1999). 
Hence, for $\beta_e (r) > 0$, $\gamma (r)$ will be $ < 1$ and the plasmoid will experience an
enhanced acceleration as discussed below (see Fig. \ref{fig:BETA}). 

Although bow shock structures generate a range of \TX\ around a plasmoid at any given
radius (see CAS08), in this paper we are only concerned with the maximum \TX\
that occurs at the apex of the bow shock. From the Rankine-Hugoniot temperature relation, this
maximum bow shock \TX(r) is
\begin{equation}
\TX\ (r) = 14 \left( \frac{\Delta V (r)}{1000} \right)^2 ~~~ MK
\label{eq:TX} 
\end{equation}
where $\Delta V (r) = \VP (r) - \VW (r)$ (measured in \kms) is the maximum (apex) relative
velocity between the plasmoid and the wind. For the ambient wind velocity (\VW) we adopt the
commonly used expression, $\VW (r) = \vinf (1~ - ~\Rstar / r)$. For $\Delta V > 0$, the plasmoid
is moving faster than the ambient wind and the bow shock is formed on the outward-facing side of
the plasmoid. Conversely, for $\Delta V < 0$, the bow shock is formed on the star-ward side of
the plasmoid since it is moving slower than the wind, a condition more likely at large wind radii.
\begin{figure}
\includegraphics[width=7.5cm,angle=180]{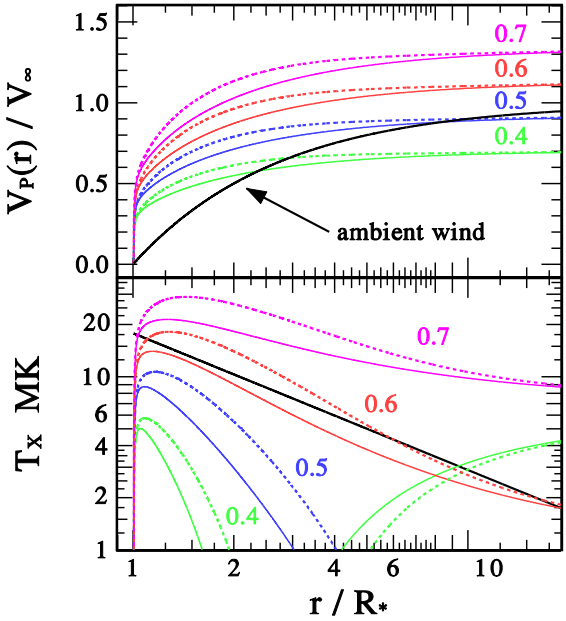}
\caption{Parameter study of the plasmoid model \VP\ (top panel) and \TX\ (bottom panel)
dependence on $\VA / \vinf = 0.4, ~0.5,~ 0.6, ~0.7$ for a fixed value of $\beta_{eo} = 1$. For
each $\VA / \vinf$, two curves are presented, for $S = 2$ (solid line) and $S = 3$ (dashed line). 
In top panel, the ambient wind velocity law is shown as a solid black line. In bottom panel, the
solid black line represents the power-law fit given by Eq. (\ref{eq:TPL}). \label{fig:VA}}
\end{figure}

We present a series of models to illustrate the dependence of \TX\ on the three basic parameters
(\VA, $\beta_{eo}$, $S$). We use a $\vinf = 2158$ \kms\ and a $\ves = 782$ \kms\ which
represent averages of the OB supergiants with measured \TX\ values.
We set $\ro = 1.005$ \Rstar\ which corresponds to an initial ambient wind velocity of 
$\sim 1/2$ times the thermal speed using a mean effective temperature of 36,000 K. 
The functional radial dependencies of \TX\ on $\VA / \vinf$, $\beta_{eo}$, and $S$ are shown in
Figures \ref{fig:VA} and \ref{fig:BETA}. As is evident in Figure \ref{fig:VA}, the ratio $\VA /
\vinf$ is the dominant parameter determining the overall magnitude of the observed \TX\
distribution. Figure \ref{fig:BETA} shows that large changes in $\beta_{eo}$ are required to 
affect the \TX\ distribution, but only below 2 \Rstar. An increase in $\beta_{eo}$ produces a rise
in the peak \TX\ and shifts it to smaller radii. The field divergence parameter, $S$, primarily
produces small vertical shifts in the \TX\ distribution.

\begin{figure}
\includegraphics[width=7cm,angle=-90]{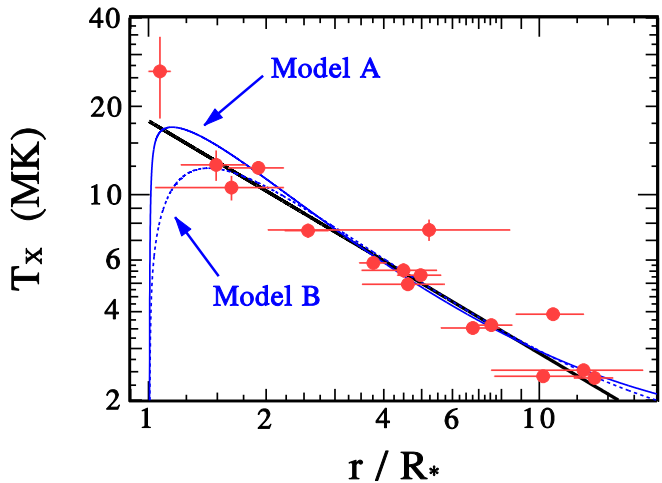}
\caption{Comparison of the two plasmoid ejection model fits to the observed \TX\ distribution of
OB supergiants; Model A (solid blue line) and Model B (dashed blue line). See Table 1 and
discussion in \S 3. The solid black line represents the power-law fit given by Eq. (\ref{eq:TPL}).
\label{fig:SGFIT}}
\end{figure}

Non-linear least-squared fit models (A \& B) are obtained by fitting the model \TX\ (Eq.
\ref{eq:TX}) to the observed OB supergiant \TX\ distribution using the supergiant \vinf\ and \ves\
averages. The model fits are shown in Figure \ref{fig:SGFIT} and best-fit
parameters are listed in Table 1. Model A uses all the data, and Model B neglects
the very high \TX\ ($\approx 26$ MK) at $\approx 1.1 \Rstar$ that is associated with \cygate.
As evident in our parameter study, both models predict a well-constrained value for $\VA /
\vinf$, emphasizing the importance of this parameter in explaining the \TX\ distribution. As for
the other two parameters, the constraints are less stringent as shown in Table 1.
The high $\beta_{eo}$ is necessary to fit the \cygate\ low radius, high \TX\ value, as
demonstrated in our $\beta_{eo}$ parameter study shown in Figure \ref{fig:BETA}.
As shown in Figure \ref{fig:SGFIT}, the two model fits (A and B) are essentially identical above
2 \Rstar, whereas, below 2 \Rstar, the models are more sensitivity to the values of $\beta_{eo}$
and $S$.
\input{tab1.tex}

From the derived value of $\VA = 0.60~ \vinf$ we can check the assumption that $\beta_{po}
<< 1$, which is fundamental to the validity of equation (\ref{eq:VP}), and we also derive a
plasmoid initial density. By writing $\beta_{po} = 2 {(a_p / \VA)}^2 = 0.02~T_6$, where
$a_p$ is the plasmoid initial thermal speed ($T_6$ has units of MK), a value of $\beta_{po} <
0.1$ requires a $T_6 < 5$. To satisfy the requirement for an enhanced ionization state in the
plasmoid as discussed earlier, all that is needed is a plasmoid temperature $> $ 0.2 MK, well
below this upper limit. 
Assuming that the detected field of $\approx 80$ G from \zoriA\ is typical for the initial
external field in all OB supergiants, we estimate from \VA\ an OB supergiant plasmoid initial
density of $4\times 10^{10}(1 + \beta_{eo})$ cm$^{-3}$ where we have used the total pressure
balance equation, ${B_p (r_o)}^2 = {B_e (r_o)}^2 (1 + \beta_{eo})$. 
Since this density is much lower than that expected for the surrounding external gas close to the
star, these plasmoids should not be visualized as clumps, but more appropriately, as isolated
magnetic rarefactions.

\section{Discussion}
We have presented a novel idea that can explain the near-surface high X-ray
temperatures [i.e., NSHIP (``near-star-high-ion-problem'') introduced by WC07], and the overall
radially-decreasing \TX\ distribution observed in OB supergiants as deduced by WC07.  By using
a mechanism suggested by CP84, self-contained diamagnetic plasmoids produced near OB stellar
surfaces can be rapidly accelerated to speeds in excess of the ambient wind terminal velocity
within 2 \Rstar. These high speeds can only occur in the CP84 theory when the ejected plasmoids
are initially magnetically dominated (i.e., $\beta_{po} <<1$). The acceleration is most strongly
dependent on the plasmoid initial \Alfven\ speed (\VA). 
Since our model fits demonstrate that a specific limit of $\VA \le 0.6 \vinf$ provides a physical
explanation of the observed radially-dependent \TX\ upper bound for all normal OB stars, it
seems that we have found a new upper limit to the initial plasmoid \Alfven\ speed for these stars.
In addition, our derived $\beta_{eo}$ and $S$ parameters provide limits on the initial
external plasma conditions. Our results demonstrate that $Chandra$ observations have the
potential to provide interesting information on the plasmoids and external plasma properties, and
provide magnetic field constraints which are particularly relevant for those OB stars for which
Zeeman measurements are not yet possible. Given the successful aspects and predictions of this
plasmoid ejection model, further observations and detailed dynamical studies of magnetic
propulsion mechanisms are clearly warranted.

\acknowledgments
We thank Vladimir Mirnov and Nicholas Murphy for helpful conversations. This work was
supported in part by awards GO2-3027A and AR8-9003A issued by the \Chandra\ X-ray
Observatory Center, and the NSF Center for Magnetic Self-Organization in Laboratory and
Astrophysical Plasmas. \Chandra\ is operated by the Smithsonian Astrophysical Observatory
under NASA contract NAS8-03060.

\end{document}

%% file: tab1.tex
\begin{deluxetable}{ccccc}
\tablecolumns{5}
\tablecaption{Plasmoid Model Best-Fit Parameters \label{tab:TABLE1}}
\tablehead{
\colhead{Model}                  & 
\colhead{$\VA / \vinf$}          & 
\colhead{$\beta_{eo}$}           & 
\colhead{$S$}                    & 
\colhead{$\chi^2 / DOF$}} 
\startdata

Model A  &  $0.61 \pm 0.01$  & $1.3 \pm 0.9$ &  $2.2 \pm 0.3$  & $17/13$  \\ 
Model B  &  $0.60 \pm 0.01$  & $0.28 \pm 0.31$ &  $2.6 \pm 0.2$  & $15/12$  \\ 

\enddata
\end{deluxetable}